# Giant Enhancement of Photoluminescence Emission in Monolayer WS$_2$ by Femtosecond Laser Irradiation

Chengbing Qin,[1,2*] Xilong Liang,[1,2] Shuangping Han,[1,2] Guofeng Zhang,[1,2] Ruiyun Chen,[1,2] Jianyong Hu,[1,2] Liantuan Xiao,[1,2*] and Suotang Jia[1,2]

[1] State Key Laboratory of Quantum Optics and Quantum Optics Devices, Institute of Laser Spectroscopy, Shanxi University, Taiyuan, Shanxi 030006, China.

[2] Collaborative Innovation Center of Extreme Optics, Shanxi University, Taiyuan, Shanxi 030006, China.

[*]Author to whom correspondence should be addressed.
Chengbing Qin, chbqin@sxu.edu.cn; Liantuan Xiao, xlt@sxu.edu.cn

**Abstract**

Monolayer transition metal dichalcogenides have emerged as promising materials for optoelectronic and nanophotonic devices. However, the low photoluminescence (PL) quantum yield (QY) hinders their various potential applications. Here we engineer and enhance the PL intensity of monolayer $WS_2$ by femtosecond laser irradiation. More than two orders of magnitude enhancement of PL intensity as compared to the as-prepared sample is determined. Furthermore, the engineering time is shortened by three orders of magnitude as compared to the improvement of PL intensity by continuous-wave laser irradiation. Based on the evolution of PL spectra, we attribute the giant PL enhancement to the conversion from trion emission to exciton, as well as the improvement of the QY when exciton and trion are localized to the new-formed defects. We have created microstructures on the monolayer $WS_2$ based on the enhancement of PL intensity, where the engineered structures can be stably stored for more than three years. This flexible approach with the feature of excellent long-term storage stability is promising for applications in information storage, display technology, and optoelectronic devices.

## 1. Introduction

Inspired by the significant advances of graphene, the transition metal dichalcogenides (TMDs) with finite bandgaps have begun to receive increasing attention in the last decade [1-3]. Unique optical and electrical properties of these TMDs, including massive excitonic effect [4,5], remarkable many-body interaction [6,7], and giant spin-valley coupling [8,9], make them a promising candidate for next-generation optoelectronic devices. However, the intrinsic defects, such as vacancies and impurities for as-prepared monolayers, generally result in the extremely low photoluminescence (PL) quantum yield (QY, ranging from 0.01% to 6% at room temperature [10]), obscuring the further development of high quantum efficiency optoelectronic devices [11]. Recently, many research efforts have been devoted to enhance the PL intensity of monolayer $WS_2$, which may be distinguished into four possibilities: (1) surface passivation with organic superacids [10,12,13], (2) energy transfer from quantum dots (QDs) [11,14-16], (3) plasmonic coupling through metallic nanostructures [17], and (4) lifting emission rate by photonic crystals [18,19].

Surface passivation is a widely known method that can tune the optical properties of two-dimensional (2D) materials. To date, the nonoxidizing organic superacid bis(trifluoromethane) sulfonamide (TFSI) [10,20], oleic acid (OA) [13] and sodium sulphide ($Na_2S$) [12] have been used to treat as-grown $WS_2$ monolayer. The total PL has been enhanced by over two orders of magnitude, yielding near-unity PL QY in monolayer TMDs. On the other hand, perovskite ($CsPbBr_3$ [16], $BA_2MA_3Pb_4I_{13}$ [15]), semiconductor (InGaN) [11], and heterostructural core/shell ($WO_3$) [14] QDs have been adopted to improve the PL intensity of WS$_2$. The mechanism is attributed to the energy transfer from QDs to TMDs, or the aggregation of excitons in WS$_2$. Other studies have sought to boost PL intensity by enhancing light-matter interaction and exciton-coupled surface plasmon polaritons. Fei *et al.* have demonstrated that the PL intensity of WS$_2$ can be improved by more than one order of magnitude in the Ag nanowire/WS$_2$/Ag film hybrid structure [17]. Very recently, over 300-fold enhancement on PL of monolayer WS$_2$ has been determined with a tailored photonic crystal structure by Jiang group [19]. This enhancement is attributed to the improvement of PL efficiency and the acceleration of spontaneous emission rate. However, the harsh nature of chemical treatment or plasmonic coupling necessitates additional fabrication steps [13]. Furthermore, contaminants and impurities are always inevitably introduced during the fabrications, which may significantly reduce the performance of optoelectronic devices based on these 2D materials [21]. Therefore, achieving high and stable PL intensity of monolayer WS$_2$, through an all-optical scheme without any chemical treatments and additional fabrication steps, is still an exploratory of fundamental research.

In this work, we report the giant and flexible enhancement of PL emission in monolayer WS$_2$ by an all-optical scheme with femtosecond laser irradiation. More than two orders of magnitude enhancement in PL intensity as compared to the as-prepared sample had been readily determined. To reveal the underlying mechanism, we performed PL spectra under different conditions. To optimize the enhancement effect, we investigated the enhancement of PL intensity as a function of irradiation time under various power densities. We also examined the stability of this enhancement by storing the processing samples in a vacuum chamber filled with nitrogen gas. At last, we demonstrated various micropatterning on monolayer $WS_2$ based on this giant, flexible, and robust PL enhancement. Benefiting from these features, our approach provides a new platform for the fabrication of information storage and nanophotonic devices based on these 2D materials.

## 2. Results and discussions

Monolayer $WS_2$, purchased from SixCarbon Technology Shenzhen, was synthesized on the $SiO_2$/Si substrate by chemical vapor deposition (CVD) [22]. The quality of the as-prepared samples has been characterized by optical microscopy and Raman spectroscopy. Their monolayer nature with a thickness of ~0.8 nm has been confirmed by atomic force microscopy, as shown in Figure S1 in the electronic supplementary materials (ESM). The main optical experiments, including femtosecond laser irradiation, PL imaging, and spectra measurements, were performed by using a home-built scanning confocal microscope, as described in the previous works [23,24], which can also be found in Figure S2 in ESM. In particular, a femtosecond laser (Mai Tai DeepSee) at the center wavelength of 780 nm with a pulse width of about 70 fs was used to irradiate and engineer the optical properties of monolayer $WS_2$. To probe the PL intensity and spectra of as-prepared and post-processed $WS_2$, a 405 nm continuous-wave (CW) laser was used and collimated with the femtosecond laser. To avoid thermal damage or further engineering, the excitation power density of the CW laser was limited to low regimes (~0.21 kW/cm$^2$). Monolayer $WS_2$ was placed on a motorized three-dimensional piezoelectric ceramic stage. The micro-patterning was achieved by moving the WS$_2$ sample with respect to the focused laser beam in a programmable way.

Very recently, some groups have engineered and promoted optical properties of monolayer MoS$_2$ and WS$_2$ by CW lasers with high power density. The PL intensities were significantly enhanced by a factor of ~10-fold [25-27]. However, the engineered times were generally in ten minutes or even hours. To further improve PL performance and significantly shorten the engineering time, we employ the femtosecond laser with high peak power density to serve as a new platform. Figure 1 presents the typical PL image after femtosecond laser irradiation. The irradiation power density was set to 2.1 MW/cm$^2$, the irradiation time was set from 30 ms to 30 s, as the solid arrows shown in Figure 1A. PL image of each irradiated spot was enlarged in Figure 1B. Some of the engineered spots present asymmetric profile. Two possible reasons are responsible for this phenomenon. The inhomogeneous quality of monolayer WS$_2$, confirming by their inhomogeneous PL intensity (Figure S3), is assumed to be the main reason. Another rational factor is the asymmetric distribution of the laser beam after focusing by the objective (Profiles of the focused laser beams have been presented in Figure S4). From Figure 1B, we can find that PL intensities of post-engineered areas are gradually enhanced before $t$=3000 ms (Figure 1Bi). Then PL intensities of the center area sharply decrease even quench to the background. These results can be well understood. The center of the laser spot had a higher power density. Thus the crystal structure of monolayer WS$_2$ was primarily damaged by femtosecond laser irradiation. To quantify the enhanced result, we define the enhancement factor (*EF*) as the ratio of the integrated

PL intensity of the post-engineered area to that of the as-prepared sample. We calculated the enhancement factor for the area of full engineered spots ($EF_1$, assuming the diameter of the largest spot to be 3 μm, as the solid line shown in the inset of Figure 1C) and the center area ($EF_2$, as the solid line shown in the inset), the results are shown in Figure 1C. The maximum enhancement factor under these conditions reaches a value up to 60 times, significantly higher than the engineered results under CW laser irradiation [21,28].

To elucidate the underlying mechanism, we measured and plotted PL spectra of engineered spots (the center area) as a function of irradiation time, as shown in Figure 2A. With the prolonging of irradiation time, PL spectra were firstly enhanced and then quenched. The redshift of the PL peak position can be clearly observed, as indicated by the dotted lines. The spectral profile of the as-prepared sample shows a broad distribution, ranging from 1.60 eV to 2.10 eV, as the top panel shown in Figure 2B. While after laser irradiation, the PL peak with the highest energy was dramatically enhanced and predominated the full spectrum, as shown in the bottom panels. All the spectral profiles can be well deconvoluted into three components, as the shadows shown in Figure 2B. According to the previous works [27,29,30], we can attribute the high and media energies to the exciton ($A^0$) and trion ($A^T$) emission. The energy difference between $A^0$ and $A^T$ is between 27 meV to 36 meV, reasonably coinciding with the trion dissociation energy reported in Ref. [30,31], as the bottom panel shown in Figure 2C. The low energy peak can be assigned to the defect-bound localized excitons ($A^D$) [29,30,32], which shows significant spectral broadening due to complex grain boundary defects and nanosize defects with different sizes [33]. The evolution of grain boundaries and nanosize defects during femtosecond laser irradiation results in the variation of its peak energy [34], as the triangle date shown in Figure 2C. Interestingly, similar evolutions are also observed for full width at half maximum (FWHM) of $A^D$, as shown in Figure S5. In contrast, peak energies and FWHM of $A^0$ and $A^T$ show little variation.

On the other aspect, the integrated PL intensities of $A^0$ and $A^T$ present dramatic change during irradiation (Figure S5). The ignorable intensity of $A^0$ of the as-prepared sample was enhanced to the dominating component after laser irradiation for more than 100 ms. To quantify the intensity evolution, we plotted the spectral weight (the ratio of the integrated PL intensity of each component to the total intensity) as the function of irradiation time, as shown in Figure 2D. Note that the weight of $A^T$ almost stands at a plateau. The weight of $A^D$ sharply decreases while that of $A^0$ rapidly increases. We speculate on two scenarios for these results. Firstly, with the prolonging of femtosecond laser irradiation, many new defects form and subsequent gas molecules (such as $O_2$ and $H_2O$) bonding to the defects convert trions to excitons by depleting electrons, which is accompanied by a remarkable enhancement of PL intensity [35]. No significant PL enhancement under the vacuum condition provides further evidence for this hypothesis (Figure S6). Another possible reason for such giant enhancement is that the quantum efficiency of quasiparticles (both excitons and trions) localized at the defects will be significantly enhanced, as suggested by Nan and Kim [35,36]. As a consequence, PL intensities of both trions and excitons are greatly improved. However, the spectral weight of trions drops slightly during irradiation, as shown in Figure 2D. Combining with the hypothesis of gas adsorption on new-formed defects as well as the ignored enhancement under vacuum condition, we suggest that the conversion from trions to excitons also plays a vital role in the PL enhancement. After irradiation for 2.0 s and more, PL intensities of all components quenched sharply (Figure S5) as the crystal structure of monolayer $WS_2$ was damaged.

The established mechanism inspires us that PL enhancement depends on the competition between the conversion from trions to excitons and the degree of crystal damage. To optimize the enhancement factor, we collected PL intensities as a function of irradiation times under different power densities on many monolayers (Figure S7), the processing results are plotted in Figure 3A. We can find that the enhancement factor reaches up to more than 200 at the power close to 8.3 MW/cm$^2$ and irradiation time of 1.0 s, as presented in Figure 3B. In our experiment, we achieved the maximum enhancement factor to be 450 (Figure S8). To guide the effective PL enhancement, we define the maximum enhancement factor ($EF_{max}$) and its reaching time ($T_{max}$), and plot these two values as a function of the irradiation power density, as shown in Figure 3C and 3D, respectively. Note that $EF_{max}$ firstly ascends with the increase of the power density, and then descends after the power density larger than 8.3 MW/cm$^2$, indicating that PL intensity cannot be further enhanced by monotonically increasing irradiation power. Unlike $EF_{max}$, the $T_{max}$ monotonically shortens with the increase of irradiation power density. And, more remarkable, $T_{max}$ is generally in sub-second. As compared to the irradiation by CW lasers [25,26,28], the engineered times have been shortened by three orders of magnitude, which significantly improves their practical applications.

Intriguingly, the ascent and descent of $EF_{max}$ as the function of the irradiation power density can be well fitted by the mono-exponential function, $EF_{max}(P) = EF_{max}(0) + A \cdot \exp[-(P-P_0)/\tau]$. Here, $P$ is the irradiation power density in MW/cm$^2$, $\tau$ represents the ascent or descent rate, determining to be 1.55 and 1.60 MW/cm$^2$ for the ascent and descent processes, respectively. Similarly, $T_{max}$ can be well fitted by mono-exponential function as well. The fitting result gives the shortening rate of 5.80 MW/cm$^2$. These values offer a quantitative description to controllably design PL intensity of monolayer WS$_2$ by femtosecond laser irradiation.

The precise control of PL intensity allows for direct and flexible micropatterns on monolayer WS$_2$. As shown in Figure 4, many desired micropatterns based on the enhancement of PL intensity have been created readily, such as English letters of "WS2" and Chinese character of "光" (meaning "light"). Note that the enhancement of PL intensity can be controlled by changing both the irradiation power density and irradiation time, here we fixed the irradiation power at 8.3 MW/cm$^2$ and irradiation time of 1.0 s for each pixel. To explore the PL enhancement, we plot the PL intensity of as-prepared and engineered area along the dashed lines shown in Figure 4A. The enhancement factor can be up to 50, as presented in Figure 4B. Furthermore, periodic arrays of microstructures (named micro-grating) can be readily created as well, as shown in Figure 4D. The corresponding enhancement profiles along with the yellow rectangle and their multi-Gauss fits have been presented in Figure 4E (PL image of the as-prepared and the relevant fitting results can be found in Figure S9 and Table S2). Almost uniform FWHM of engineered lines indicates the precision control of PL enhancement by femtosecond laser irradiation. This flexible micropatterning enables promising applications in optical data storage, display technology, and micro-optoelectronic devices. To further explore the promise of this technique, we characterized the stability and reliability of micropatterning after irradiation. The processing sample was stored in an improvised vacuum chamber filled with nitrogen gas. All the optical experiments were performed at the ambient atmosphere without any further protection. Figure S10 presents the corresponding PL image and PL intensity profiles along the marked lines at different time points after laser irradiation. The evolution of relevant PL enhancement factors is illustrated in Figure 4F. From which we can find that the enhancements are reasonably stable in three years. The small variation possibly originates from the change of the parameters of the home-built experimental setup (such as the excitation power and focusing of the objective) and the unavoidable degradation of monolayer $WS_2$ (such as pollution by the dust and water vapor). This result further explores the promising applications of our scheme in the information storage and related optical devices.

## 3. Conclusion

In conclusion, we have shown that the PL intensity of monolayer $WS_2$ can be enhanced by a factor of more than two orders of magnitude by femtosecond laser irradiation under optimized parameters, as compared to the as-prepared sample. The maximum enhancement is 450 times in our experiment, and the irradiation time is shortened by three orders of magnitude, as compared with the irradiation by CW lasers.

This giant enhancement is attributed to not only the improvement of QY of exciton and trion when they are localized to the laser-induced defects but also the conversion from trion to exciton. In particular, this PL enhancement is robust, where the enhanced PL intensity can be stability preserved for more than three years. This giant, flexible, and robust PL enhancement allows us to create micropatterning on monolayer TMDs, which has promising applications in information storage and nano-optoelectronic devices. The enhanced PL QY also enables the development of new high-performance light-related devices, such as light-emitting diodes, lasers, and solar cells.

**Electronic supplementary materials**

Supplementary materials, i.e., optical characterization of the as-prepared sample, the schematic diagram of the optical setup, profiles of the focused laser beam, the time evolution of the integrated PL intensity and FWHM, PL evolution under a vacuum condition, power-dependent PL enhancement, the maximum PL enhanced result, characterization of the micro-grating, as well as the robust of the PL enhancement, are available in the online version.

**Acknowledgment**

This work is supported by the National Key Research and Development Program of China (Grant No. 2017YFA0304203), Natural Science Foundation of China (Nos. 91950109, 61875109, 61527824, and 61675119), Natural Science Foundation of Shanxi Province (No. 201901D111010(ZD)), PCSIRT (No. IRT_17R70), 1331KSC, PTIT, and Postgraduate education Innovation project of Shanxi Province (2019SY052 and 2020BY022).

**References**

1. H. Zeng, X. Cui. An optical spectroscopic study on two-dimensional group-VI transition metal dichalcogenides. *Chem. Soc. Rev.* 44, 2629 (2015).
2. Q. H. Wang, K. Kalantar-Zadeh, A. Kis, J. N. Coleman, M. S. Strano. Electronics and optoelectronics of two-dimensional transition metal dichalcogenides. *Nat. Nanotechnol.* 7, 699 (2012).
3. X. Xu, W. Yao, D. Xiao, T. F. Heinz. Spin and pseudospins in layered transition metal dichalcogenides. *Nat. Phys.* 10, 343 (2014).

4. C. Qin, Y. Gao, Z. Qiao, L. Xiao, S. Jia. Atomic-Layered MoS$_2$ as a Tunable Optical Platform. *Adv. Opt. Mater.* 4, 1429 (2016).

5. K. F. Mak, C. Lee, J. Hone, J. Shan, T. F. Heinz. Atomically Thin MoS$_2$: A New Direct-Gap Semiconductor. *Phys. Rev. Lett.* 105, 136805 (2010).

6. D. Y. Qiu, F. H. da Jornada, S. G. Louie. Optical Spectrum of MoS$_2$: Many-Body Effects and Diversity of Exciton States. *Phys. Rev. Lett.* 111, 216805 (2013).

7. K. F. Mak, K. He, C. Lee, G. H. Lee, J. Hone, T. F. Heinz, J. Shan. Tightly bound trions in monolayer MoS$_2$. *Nat. Mater.* 12, 207 (2013).

8. H. Zeng, J. Dai, W. Yao, D. Xiao, X. Cui. Valley polarization in MoS$_2$ monolayers by optical pumping. *Nat. Nanotechnol.* 7, 490 (2012).

9. D. Xiao, G.-B. Liu, W. Feng, X. Xu, W. Yao. Coupled Spin and Valley Physics in Monolayers of MoS$_2$ and Other Group-VI Dichalcogenides. *Phys. Rev. Lett.* 108, 196802 (2012).

10. H. Kim, D. H. Lien, M. Amani, J. W. Ager, A. Javey. Highly Stable Near-Unity Photoluminescence Yield in Monolayer MoS$_2$ by Fluoropolymer Encapsulation and Superacid Treatment. *ACS Nano* 11, 5179 (2017).

11. G. Cheng, B. Li, C. Zhao, Z. Jin, H. Li, K. M. Lau, J. Wang. Exciton aggregation induced photoluminescence enhancement of monolayer WS$_2$. *Appl. Phys. Lett.* 114, 232101 (2019).

12. H. Yao, L. Liu, Z. Wang, H. Li, L. Chen, M. E. Pam, W. Chen, H. Y. Yang, W. Zhang, Y. Shi. Significant photoluminescence enhancement in WS$_2$ monolayers through Na2S treatment. *Nanoscale* 10, 6105 (2018).

13. A. O. A. Tanoh, J. Alexander-Webber, J. Xiao, G. Delport, C. A. Williams, H. Bretscher, N. Gauriot, J. Allardice, R. Pandya, Y. Fan, et al. Enhancing Photoluminescence and Mobilities in WS$_2$ Monolayers with Oleic Acid Ligands. *Nano Lett.* 19, 6299 (2019).

14. C. Zou, M. Chen, X. Luo, H. Zhou, T. Yu, C. Yuan. Enhanced photoluminescence of WS$_2$/WO$_3$ heterostructural QDs. *J. Alloys Compd.* 834, 155066 (2020).

15. A. Yang, J. C. Blancon, W. Jiang, H. Zhang, J. Wong, E. Yan, Y. R. Lin, J. Crochet, M. G. Kanatzidis, D. Jariwala, et al. Giant Enhancement of Photoluminescence

Emission in WS$_2$-Two-Dimensional Perovskite Heterostructures. *Nano Lett.* 19, 4852 (2019).

16. Y. Liu, H. Li, X. Zheng, X. Cheng, T. Jiang. Giant photoluminescence enhancement in monolayer WS$_2$ by energy transfer from CsPbBr$_3$ quantum dots. *Opt. Mater. Express* 7, 1327 (2017).

17. F. Cheng, A. D. Johnson, Y. Tsai, P.-H. Su, S. Hu, J. G. Ekerdt, C.-K. Shih. Enhanced Photoluminescence of Monolayer WS$_2$ on Ag Films and Nanowire–WS$_2$–Film Composites. *ACS Photonics* 4, 1421 (2017).

18. J. Wang, H. Li, Y. Ma, M. Zhao, W. Liu, B. Wang, S. Wu, X. Liu, L. Shi, T. Jiang, et al. Routing valley exciton emission of a WS$_2$ monolayer via delocalized Bloch modes of in-plane inversion-symmetry-broken photonic crystal slabs. *Light Sci. Appl.* 9, 148 (2020).

19. H. Li, J. Wang, Y. Ma, J. Chu, X. a. Cheng, L. Shi, T. Jiang. Enhanced directional emission of monolayer tungsten disulfide (WS$_2$) with robust linear polarization via one-dimensional photonic crystal (PhC) slab. *Nanophotonics* (2020). *https://doi.org/10.1515/nanoph-2020-0294*

20. M. Amani, D. H. Lien, D. Kiriya, J. Xiao, A. Azcatl, J. Noh, S. R. Madhvapathy, R. Addou, K. C. Santosh, M. Dubey. Near-unity photoluminescence quantum yield in MoS$_2$. *Science* 350, 1065 (2015).

21. C. Yang, Y. Gao, C. Qin, X. Liang, S. Han, G. Zhang, R. Chen, J. Hu, L. Xiao, S. Jia. All-Optical Reversible Manipulation of Exciton and Trion Emissions in Monolayer WS$_2$. *Nanomaterials* 10, 23 (2020).

22. D. Zhou, H. Shu, C. Hu, L. Jiang, P. Liang, X. Chen. Unveiling the Growth Mechanism of MoS$_2$ with Chemical Vapor Deposition: From 2D Planar Nucleation to Self-Seeding Nucleation. *Cryst. Growth Des.* 18, 1012 (2018).

23. W. He, C. Qin, Z. Qiao, G. Zhang, L. Xiao, S. Jia. Two fluorescence lifetime components reveal the photoreduction dynamics of monolayer graphene oxide. *Carbon* 109, 264 (2016).

24. Z. Qiao, C. Qin, W. He, Y. Gong, G. Zhang, R. Chen, Y. Gao, L. Xiao, S. Jia. Versatile and scalable micropatterns on graphene oxide films based on laser induced fluorescence quenching effect. *Opt Express* 25, 31025 (2017).

25. H. Ardekani, R. Younts, Y. Yu, L. Cao, K. Gundogdu. Reversible Photoluminescence Tuning by Defect Passivation via Laser Irradiation on Aged Monolayer $MoS_2$. *ACS Appl. Mater. Interfaces* 11, 38240 (2019).

26. H. M. Oh, G. H. Han, H. Kim, J. J. Bae, M. S. Jeong, Y. H. Lee. Photochemical Reaction in Monolayer $MoS_2$ via Correlated Photoluminescence, Raman Spectroscopy, and Atomic Force Microscopy. *ACS Nano* 10, 5230 (2016).

27. P. K. Chow, R. B. Jacobs-Gedrim, J. Gao, T.-M. Lu, B. Yu, H. Terrones, N. Koratkar. Defect-Induced Photoluminescence in Monolayer Semiconducting Transition Metal Dichalcogenides. *ACS Nano* 9, 1520 (2015).

28. C. Qin, Y. Gao, L. Zhang, X. Liang, W. He, G. Zhang, R. Chen, J. Hu, L. Xiao, S. Jia. Flexible engineering of light emission in monolayer $MoS_2$ via direct laser writing for multimode optical recording. *AIP Adv.* 10, 045230 (2020).

29. Y. Lee, S. J. Yun, Y. Kim, M. S. Kim, G. H. Han, A. K. Sood, J. Kim. Near-field spectral mapping of individual exciton complexes of monolayer $WS_2$ correlated with local defects and charge population. *Nanoscale* 9, 2272 (2017).

30. V. Carozo, Y. X. Wang, K. Fujisawa, B. R. Carvalho, A. McCreary, S. M. Feng, Z. Lin, C. J. Zhou, N. Perea-Lopez, A. L. Elias, et al. Optical identification of sulfur vacancies: Bound excitons at the edges of monolayer tungsten disulfide. *Sci. Adv.* 3, 9 (2017).

31. Y. Lee, G. Ghimire, S. Roy, Y. Kim, C. Seo, A. K. Sood, J. I. Jang, J. Kim. Impeding Exciton-Exciton Annihilation in Monolayer $WS_2$ by Laser Irradiation. *ACS Photonics* 5, 2904 (2018).

32. J. Hong, M. Wang, J. Jiang, P. Zheng, H. Zheng, L. Zheng, D. Huo, Z. Wu, Z. Ni, Y. Zhang. Optoelectronic performance of multilayer $WSe_2$ transistors enhanced by defect engineering. *Appl. Phys. Express* 13, 061004 (2020).

33. Y. Lee, S. Park, H. Kim, G. H. Han, Y. H. Lee, J. Kim. Characterization of the structural defects in CVD-grown monolayered $MoS_2$ using near-field photoluminescence imaging. *Nanoscale* 7, 11909 (2015).

34. S. Tongay, J. Suh, C. Ataca, W. Fan, A. Luce, J. S. Kang, J. Liu, C. Ko, R.

Raghunathanan, J. Zhou, et al. Defects activated photoluminescence in two-dimensional semiconductors: interplay between bound, charged, and free excitons. *Sci Rep* 3, 2657 (2013).

35. H. J. Kim, Y. J. Yun, S. N. Yi, S. K. Chang, D. H. Ha. Changes in the Photoluminescence of Monolayer and Bilayer Molybdenum Disulfide during Laser Irradiation. *ACS Omega* 5, 7903 (2020).

36. H. Nan, Z. Wang, W. Wang, Z. Liang, Y. Lu, Q. Chen, D. He, P. Tan, F. Miao, X. Wang. Strong Photoluminescence Enhancement of $MoS_2$ through Defect Engineering and Oxygen Bonding. *ACS Nano* 8, 5738 (2014).

**Figures and captions:**

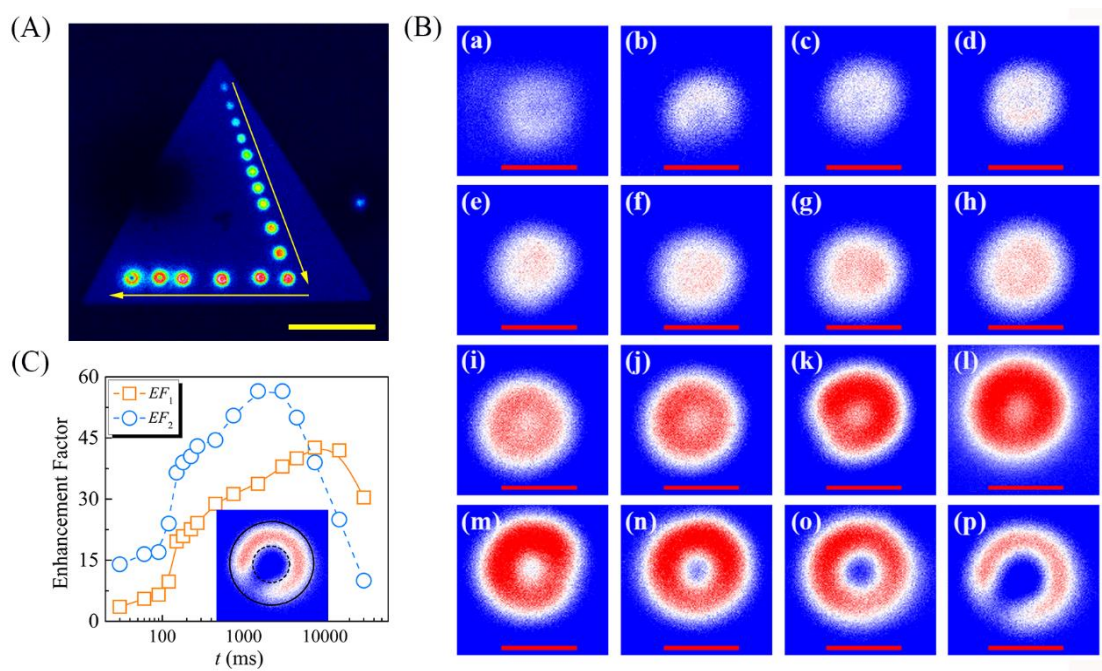

Figure 1. (A) Photoluminescence (PL) image of monolayer $WS_2$ of as-prepared and post-engineered by femtosecond laser with the increase of irradiation time, indicated by the solid arrows. Scale bar: 10 μm. (B) Enlarged image of each engineered spot. Scale bar: 2 μm. The irradiation times from (a) to (p) were 30, 60, 90, 120, 150, 180, 225, 270, 450, 750, 1500, 3000, 4500, 7500, 15000, and 30000 ms, respectively. The irradiation power density of the femtosecond laser was 2.1 MW/cm². The excitation laser for PL image was 405 nm CW laser with the power density of 0.21 kW/cm². (C) The enhancement factor of the full ($EF_1$) and the center ($EF_2$) areas of the engineered spots (as the solid and dashed circles shown in the inset) as a function of irradiation time (in log scale).

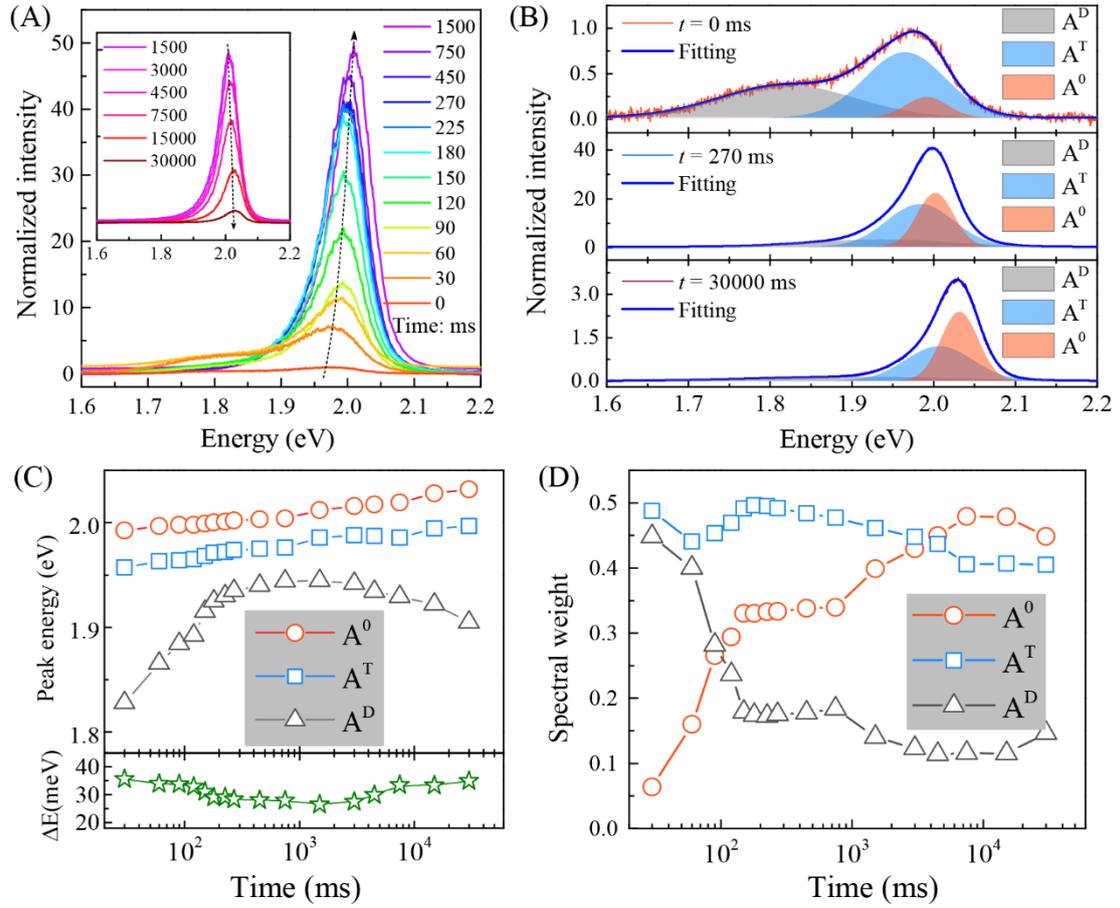

Figure 2. Time evolution of PL spectra of monolayer $WS_2$ under femtosecond laser irradiation. (A) Measured PL spectra as a function of laser irradiation time for the enhancement stage (from 30 to 1500 ms) and the quenching stage (from 1500 to 30000 ms in the inset), respectively. All the PL intensity is normalized to that of the as-prepared sample. (B) Representative PL spectra under three irradiation times, all the spectral profiles are deconvoluted into three peaks (exciton, $A^0$; trion, $A^T$; and defect-bound localized exciton, $A^D$) using Gauss function. (C) Peak energies and (D) the spectral weight of three components as a function of laser irradiation time (in log scale), respectively. The bottom panel presents the trion dissociation energy, $\Delta(E_{A0}-E_{AT})$, varied as laser irradiation time.

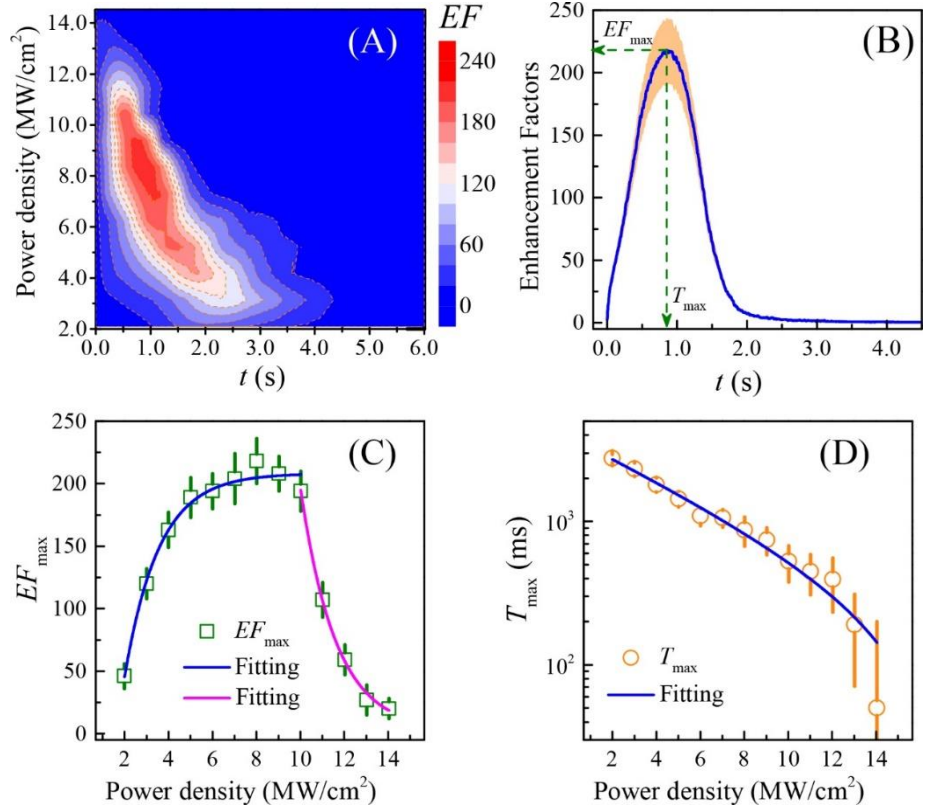

Figure 3. (A) The contour map of the power-dependent PL enhancement factor (*EF*) trajectory (enhancement factor as a function of irradiation time). (B) PL enhancement factor trajectory line-outs at the power of 8.0 MW/cm$^2$. The definitions of maximum enhancement factor ($EF_{max}$) and its reaching time ($T_{max}$) are indicated. The color bar denotes the corresponding standard deviations. (C) $EF_{max}$ and (D) $T_{max}$ as a function of the power density. The solid lines are the fitting curves according to the exponential function.

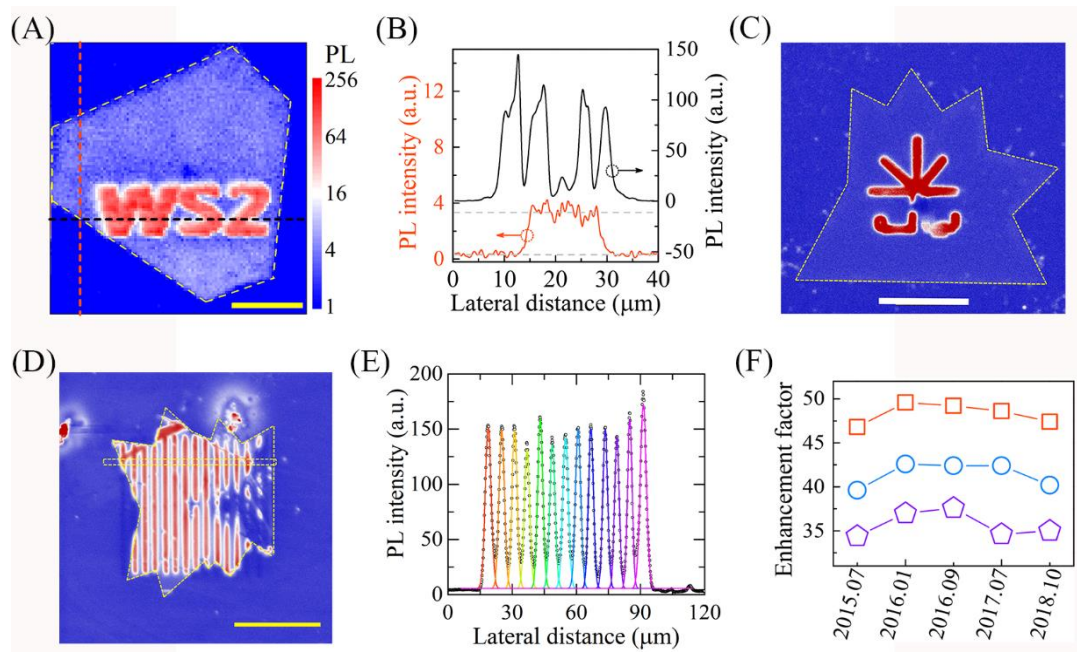

Figure 4. Patterning of nanostructures on monolayer WS$_2$ based on the enhancement of PL intensity. PL image of engineered letters "WS2" (A), the Chinese character "光" (C), and micro-grating (D), respectively. The irradiation power density of the femtosecond laser was 8.3 MW/cm$^2$; the irradiation time was 1 s. The excitation laser for PL image was 405 nm CW laser with the power density of 0.21 kW/cm$^2$. Scale bar: 10 μm in A, 20 μm in C, 50 μm in D, respectively. (B) PL intensities along the marked lines in A. (E) PL intensity of the selected area in D. PL intensity profile was fitted by multi-Gauss function, as the solid lines shown in the figure. (F) Evolution of PL enhancement factor at different time points. The squares, circles, and pentagons denote the three engineered spots with the power density of 8.3 MW/cm$^2$ and irradiation time of 800 ms, 1000 ms, and 1200 ms, respectively.